\documentclass[aps,pra, 10pt,superscriptaddress,twocolumn,floatfix]{revtex4-2}
\usepackage{graphicx,psfrag,amsmath,amssymb,amsfonts,latexsym,color,epsf,dcolumn,graphpap, mathtools}
\usepackage{caption}
\usepackage{xfrac}
\usepackage{xcolor}
\usepackage{subfig}
\usepackage{lipsum}
\usepackage{comment}
\usepackage[colorlinks,citecolor=magenta]{hyperref}

\usepackage{cancel}
\allowdisplaybreaks

\definecolor{red}{rgb}{1,0,0}
\definecolor{blue}{rgb}{0,0,1}
\definecolor{skyblue}{rgb}{0,0,.5}
\definecolor{green}{rgb}{0,1,0}
\definecolor{orange}{cmyk}{0,.4,1,0}
\definecolor{darklavender}{rgb}{0.45, 0.31, 0.59}
\definecolor{darkorchid}{rgb}{0.6, 0.2, 0.8}
\definecolor{dimgray}{rgb}{0.41, 0.41, 0.41}
\definecolor{dogwoodrose}{rgb}{0.84, 0.09, 0.41}

\def \C {{\scriptscriptstyle \rm C}}
\def \B {{\scriptscriptstyle \rm QB}}

\begin{document}
\title{Geometric phase in dissipative quantum batteries}

\author{Camila Cristiano}
\affiliation{Basque Center for applied mathematics, Alameda de Mazarredo 14, 48009 Bilbao, Spain}
\author{Ludmila Viotti}
\affiliation{The Abdus Salam International Center for Theoretical Physics, Strada Costiera 11, 34151 Trieste, Italy}
\author{Paula I.~Villar}
\affiliation{Departamento de F\'\i sica {\it Juan Jos\'e
 Giambiagi}, FCEyN UBA, Facultad de Ciencias Exactas y Naturales, Ciudad Universitaria, Pabell\' on I, 1428 Buenos Aires, Argentina }
 \affiliation{IFIBA CONICET-UBA, Facultad de Ciencias Exactas y Naturales,
 Ciudad Universitaria, Pabell\' on I, 1428 Buenos Aires, Argentina }

\begin{abstract}
We study the geometric phase accumulated during non-adiabatic charging of different driven open quantum systems serving as quantum battery models. We provide a full numerical analysis of dynamics under different type of noises typically reported in superconducting circuits implementations. We complement the study with analytic results derived in the limiting case of no noise (i.e. isolated systems). We compute the non-unitary geometric phase acquired by the quantum batteries during the transition and show that there is a direct relation between the accumulated geometric phase and the integral of the stored energy during the transition. Finally, we perform the same analysis on a bipartite quantum battery that relies on a dephased charger and found similar results. Our theoretical findings 
are within experimental reach using state-of-the-art techniques.
\end{abstract}

\noindent 

\maketitle
\section{Introduction}

Driven quantum systems play a central role in quantum information processing, serving as a paradigmatic model that captures the essential physics of a wide range of systems. Initially introduced in the study of spin dynamics and atomic collisions, the concept of a driven two-level system has been extended to artificial mesoscopic platforms, such as semiconductor quantum dots and superconducting circuits. 
Solid-state implementations of these systems have attracted considerable attention, as they provide a unique opportunity to observe fundamental quantum phenomena at macroscopic scales and represent promising candidates for qubit realization in emerging quantum technologies.
 
With the advancement of quantum technologies, the thermodynamic behavior of quantum systems has become a subject of growing interest. In particular, both theoretical and experimental studies have investigated how engineered quantum systems can act as energy reservoirs capable of supplying this energy on demand. This behavior naturally leads to the idea of quantum batteries~\cite{Campaioli2018,Bhattacharjee,Campaioli2023}, and to the question of whether the performance of energy-storing devices in terms of energy transfer and charging power could be improved by exploiting quantum features. It has been stated that entanglement between multiple quantum batteries leads to a quantum advantage, as it increases the collective charging speed beyond what would be possible if each battery was charged independently~\cite{gyhm,shi}. This approach suggests that quantum batteries may one day offer more efficient and faster charging energy storage solutions.

While significant attention has been devoted to characterizing quantum devices and assessing their potential advantages over classical counterparts, theoretical efforts have also focused on identifying viable implementations. In this context, various models of quantum batteries have been proposed, primarily differing in their charging mechanisms. One main approach involves unitary energy transfer between a charger and a battery, both treated as quantum systems~\cite{Andolina2018, Crescente2022}. This case has been discussed, in particular, for arrays of artificial atoms~\cite{Campaioli2017, Le2018, Rossini2020, Rosa2020, Quach2020, Gyhm2022} and both cavity and circuit quantum electrodynamics~\cite{Delmonte2021,Seah2021}. In a second approach, the charging is achieved by the action of a classical external drive~\cite{Zhang2019, Crescente2020_v2}. 
In~\cite{Crescente2020_v2} authors analyze a driven quantum battery subjected to an AC field. 
The miniaturization technology of integrated circuits makes it an ideal architecture for implementing quantum batteries. In this promising architecture, driven quantum systems have been proposed as a model to capture the essential physics of quantum batteries.

After the seminal works, constrained to perfectly isolated setups, quantum batteries have also been considered as open systems. In such approach, the battery, the charger, or both, are coupled to a reservoir. The usual scenarios involve depletion of quantum resources. The implementation of a quantum battery in practice must address the challenge of environmental interactions, as protecting against energy leakage and decoherence is crucial for the successful realization of such devices. In~\cite{Tabesh} authors have focused on the role of common reservoirs with non-Markovian and Markovian dynamics.
Charger-mediated quantum batteries coupled to open driven chargers were addressed~\cite{Farina19, Shastri}, and the effect on the energy fluctuation along the charging process of initial state preparation was explored ~\cite{Crescente2020}.

Likewise, actual experimental progress has also been achieved in the field. The performance of the IBM Armonk single qubit was investigated in terms of energy storage and charging time characterizing the classical external drive required~\cite{Gemme1}. 
Sequential and continuous charging protocols in a qutrit have been reported and tested in an IBM quantum device~\cite{Gemme}, and  the charging and self-discharging of a superconducting quantum battery has been characterized~\cite{hu2022optimal}. The device consisted in a driven quantum system: a transmon qutrit that was externally driven by two microwave pulses.
Also open quantum batteries have been experimentally addressed. In~\cite{Dou2018} a three-level quantum battery is proposed, which consisted of a three-level system driven by two external optical or microwave fields, that realizes the energy transfer between the ground state and the maximum excited state. Similarly, an implementation scheme of a QB was proposed on a superconducting circuit composed by N coupled transmon qubits and a one-dimensional transmission line resonator~\cite{Dou2022}.

The existence of a geometric phase (GP) acquired by the state of a quantum system was stated on theoretical grounds by Berry, in the context of adiabatic and cyclic unitary evolution~\cite{Berry}. Later, the concept has been generalized to non-adiabatic (while still cyclic), non-cyclic and even non-unitary evolution~\cite{Aharonov,Samuel,Sjoqvist2000,Wilczek,Anandan,Singh,Tong, Wu}. GPs have become not only a fruitful course of investigation of fundamental features of a quantum mechanics but also of technological interest. Depending in a non-trivial way on the path traversed by the state of the system in the physical-states space, they can exhibit high sensitivity to certain environment or parametric changes, remaining robust against others. As a consequence, GPs can constitute highly effective quantum sensors.
For example, a velocity-dependent correction to the accumulated GP was reported for an NV-center at a fixed distance of a coated Si disk mounted on a turntable, which could serve as indirect detection of Casimir friction force~\cite{Npj}. 
The appearance of a vacuum-induced Berry phase in an artificial transmon atom, was demonstrated in~\cite{Pechal} and  the robustness of this GP was explained in~\cite{Viotti}.
GPs have been observed in superconducting circuits where the resonator and the qutrit energy levels are dispersively coupled, and where five qubits are controllably coupled to a resonator~\cite{Zhang2018}. In this last scenario, a quantum gate protocol based on this GP is reported. Further experimental observation of the GPs can be found in \cite{Song2017, cucchietti}.
The combination of sensitivity and robustness makes GP-based quantum sensors particularly promising in metrology and fundamental physics research, where detecting minute shifts or fields with minimal error is crucial. 

In this framework, we propose the study of the GP accumulated by the state of a 
driven-dissipative system which operates as an open QB, and address whether it is possible to infer properties of physical parameters of the QB from it. For a general answer, we explore both the dynamics of the charging process and the accumulated GP for several models of quantum batteries. The article is organized as follows. In Sec.\ref{modelo}, we present the general theoretical description of an open quantum battery and the definition of the GP for non-unitary evolutions. 
Afterwards, we turn into specific models. In Sec.\ref{2L}, we study a two-level QB charged by a classical field. In the limiting case of unitary evolution, we analytically compute the GP, and  show that it is directly related to the energy stored in the battery, due to the vanishing of the so-called Pancharatnam phase. Then, we use the unitary case as a benchmark and present the GP acquired by the battery when the charging process is carried out under noise and dissipation. In Sec.~\ref{3L}, we extend the analysis to a three-level QB  and compute the GP for an open three-level system. Again, we show that in the limit of no environment (i.e. for the close system) there is a simple relation between the accumulated GP and the stored energy. Moreover, this relation is found to be resilient to a particular type of noise.
In Sec.~\ref{charger}, we consider a model in which the charger is a quantum device, rendering the battery-charger system a bipartite quantum system, and compute the GP of the dissipative QB. Finally, we resume our conclusions in Sec.~\ref{conclusiones}.

\section{ General setup}
\label{modelo}
\subsubsection{Quantum Batteries}
A quantum battery (QB) is an energy storage device that can be modeled through a $d$-level system as
\begin{equation}
    \hat{H}_{\B}= \sum_{n = 1}^d \omega_ n |n\rangle \langle n| 
    \label{QB}
\end{equation}
where we assume $\omega_n < \omega_{n+1}$ non-degenerated energy levels and take $\hbar = 1$ hereafter~\cite{Alicki}. The storage device (i.e. the QB properly) is further coupled to a charger which is responsible of inducing transitions among energy levels, and can be modeled either as an auxiliary quantum system, or as a classical driving. If the battery is not perfectly protected from the outside, or if it is coupled to a quantum charger that interacts with the external environment, it becomes an open quantum battery (OQB), with the full battery-charger-environment scenario described by
\begin{equation}
    \hat{H}=  \hat{H}_{o} + \hat{H}_{\rm e} + \hat{H}_{\rm int}.
    \label{OQB}
\end{equation}
Here $\hat{H}_{o}$ accounts for the battery + charger dynamics, and $\hat{H}_{\rm int}$ defines the interaction with the environment, of internal dynamics ruled by $\hat{H}_{\rm e}$. 

After tracing out the environmental degrees of freedom, the evolution of the OQB is given by a so-called master equation. The coupling with the environment is crucial, as it intrinsically deviates the dynamics from those observed in the isolated case, introducing decoherence and energy relaxation effects that tend to deteriorate the resourceful active states, leading them to either partially or fully passive states.
As a consequence, in most cases, the environmentally induced effects contribute to shortening how long the battery can keep its charge. However, contrary to the most usual results, the use of a shared reservoir for a QB coupled to a quantum charger can establish an optimal condition where non-reciprocity improves charging efficiency and increases energy storage in the battery~\cite{Ahmadi}. 

The possibility of implementing QBs has been theorized in a variety of physical systems. Superconducting circuits are one of the architectures that have gained the most relevance in recent years due to their great potential for the development of new technologies~\cite{santos2019stable}, with transmonic batteries charged by the action of an external classical time-dependent field, a recurrent choice. 
While some authors considered OQB beyond the markovian approximation involved in the Linblad equation~\cite{Kamin2020, PhysRevA_noise},  with a quantum circuit setup in mind, we will take into account the environment through the phenomenological Lindblad-type master equation~\cite{Peterer, Blais2021}:
\begin{equation}
    \dot{\rho}(t) = -i\,[\hat{H}_o, \rho(t)] + \gamma\,\mathcal{L}_{\text{rel}}[\rho(t)] + \eta\,\mathcal{L}_{\text{dep}}[\rho(t)].
    \label{eq:Lindblad}
\end{equation}
Herein, the first term describes the unitary evolution of the QB and charger device, ${\cal L}_{\rm rel}[\rho]= \sum_k \gamma_{k} \,(\hat{L}_{{\rm r}, k}\,\rho\, \hat{L}_{{\rm r}, k}^{\dagger}-\sfrac{1}{2}\,\{\hat{L}_{{\rm r}, k}^{\dagger} \hat{L}_{{\rm r}, k},\rho\})$, where the operators $\hat{L}_{{\rm r}, k}$ generate incoherent transitions at rates $\gamma_k$ that are related to the atom-environment coupling strengths at atomic frequencies; while ${\cal L}_{\rm dep}[\rho]= \sum_k\eta_k \,(\hat{L}_{{\rm d}, k}\,\rho \hat{L}_{{\rm d}, k}^{\dagger}-\sfrac{1}{2}\,\{\hat{L}_{{\rm d}, k}^{\dagger} \hat{L}_{{\rm d}, k},\rho\})$, where $\eta_k$ are the pure dephasing rates caused, for example, by fluctuations of parameters controlling their transition frequency and by dispersive coupling to other degrees of freedom in their environment. 

In all cases listed below, we shall consider the dynamics of the charging process from a less energetic state, to a state with greater energy, ideally the ground state transitioning to an excited one.
We shall define the energy stored in the QB at time $t$ as the difference:
\begin{eqnarray}
    \mathcal{E}(t)   =  {\rm tr}(\hat{H}_{\B}\,\rho(t)) - {\rm tr}(\hat{H}_{\B}\,\rho(0)).
    \label{eqn:E_Almacenada_Definicion}
\end{eqnarray}

\subsubsection{Geometric phase}
The GP accumulated by the state of a quantum system depends solely on the path traced in the ray space, this is, the space of physical states. In terms of Hilbert space vectors, it can be written as
\begin{equation}
    \Phi_g(t) = \arg\langle\psi(0)|\psi(t)\rangle - {\rm Im}\int_0^t\,dt'\,\langle\psi(t')|\dot\psi(t')\rangle.
    \label{eq:GP}
\end{equation}
When the battery is isolated and the dynamics is unitary (but otherwise general), $|\psi(t)\rangle$ represents the state of the system at time $t$ \cite{Mukunda}. Under these circumstances, we will refer to the GP as $\Phi_g^u$, and the two terms in Eq.(\ref{eq:GP}) acquire a clear interpretation: the first term is the relative phase of the final state with respect to the initial one as determined by Pancharatnam criterion, while the second term is the subtraction of the {\it dynamic} phase out from the total phase.
On the other hand, a proper generalization of a GP which is well defined to cover non-unitary evolutions was put forward on \cite{Tong}. Conveniently, as long as the initial state is pure, the generalized GP  applying to open evolutions is also given by Eq. (\ref{eq:GP}). In such case $|\psi(t)\rangle = |\psi_+(t)\rangle$ represents the eigenstate of the state density matrix that coincides with the initial state at time $t=0$, this is, the eigenstate of $\rho(t)$ with associated eigenstate $\epsilon_+(t)$ such that $\epsilon_+ (0) = 1$. It is straightforward to see that this generalized GP reduces to the unitary result if the evolution is indeed unitary, as the density matrix eigenstate $|\psi_+(t)\rangle\equiv|\psi(t)\rangle$ coincides with the pure state of the system. Moreover, the expression also reduces to all the well-known less general results upon satisfying the specific conditions assumed for each of these. In particular, Berry's phase is recovered for cyclic adiabatic evolutions.
\\
As most experimental efforts report unitary GPs, we shall consider the non-unitary GP as an indicator revealing the range of validity of unitary results.
To do so, we can always define the deviation from the unitary result $\delta\Phi_g$ as
\begin{equation}
     \delta\Phi_g = \Phi_g - \Phi_g^u.
     \label{eq:faseabierta}
\end{equation}
Whenever the deviation from the unitary result surpasses a certain predefined value, we shall assume the effect of the environment is too strong for the unitary GP value to be expected.
These corrections have also been reported to be useful for tracing information of the system \cite{Npj}.


\begin{figure*}[t]
 \begin{center}
\includegraphics[width=0.8\linewidth]{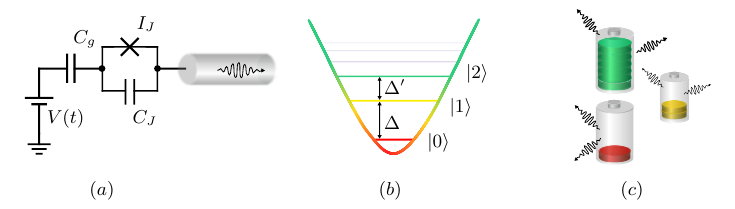}
 \caption{ (a) Scheme of a superconducting trasmon circuit capacitively coupled to an external drive and an external environment which is modeled with a semi-infinite waveguide. (b) Scheme of the energy levels of the transmon. We'll restrict to either the first two or the first three levels to model a quantum battery (c).
 }
 \label{fig:circuito}
 \end{center}
 \end{figure*}

\section{Two-level state quantum battery}
\label{2L}

We begin by considering the minimal model of a QB consisting of two levels which can be implemented, for example, in the lowest energy levels of a transmon as sketched in Fig.~\ref{fig:circuito} (with $\Delta'\gg\Delta$). The Hamiltonian of the QB in Eq.~\eqref{QB} reads, when specified for this model

\begin{equation}
  \hat{H}_{\B}^{(2)} = \omega_0|0\rangle \langle0|  + \omega_1|1\rangle \langle 1|,
  \label{eqn:2LS_H_Q}
\end{equation}
where $|0\rangle, |1\rangle$ are the two possible energy states and we define $\Delta=\omega_1-\omega_0$, the energy gap between levels. Charging a QB involves transitioning it from the ground state to the first excited state. Here, we take this process to be induced by a classical control driving described by the time-dependent interaction 

\begin{equation}
  \hat{V}^{(2)}(t) = gf(t)\cos\bigl(\Omega t \bigr) \Bigl (|0\rangle \langle 1| + |1\rangle \langle 0| \Bigr ),
  \label{eqn:2LS_H_int}
\end{equation}
where $f(t)$ is an adimensional time-dependent function $|f(t)|\leq 1$ modulated by a cosine of tuneable frequency $\Omega$, and $g$ is a constant factor setting the coupling intensity between the QB and the external classical field. Therefore, the total Hamiltonian reads

\begin{equation}
  \hat{H}_o^{(2)}(t) = \hat{H}_{\B}^{(2)} + \hat{V}^{(2)}(t).
  \label{eqn:2LS_H_tot}
\end{equation}
The model was tested in IBM-quantum devices and represents the simplest experimental version of a quantum battery~\cite{Gemme}.
Typically, experimental realizations with transmon devices consider $g/\Delta < 1$~\cite{Krantz}. In such conditions, in order to achieve a complete charging of the QB (i.e. a perfect population transition), the resonant condition $\Omega=\Delta$ is due.
We further consider a pure initial state $|\psi(0)\rangle$ of the form

\begin{equation}
    |\psi(0) \rangle = \sqrt{a}|0\rangle + \sqrt{1-a}e^{i \phi}|1\rangle 
    \label{eqn:2LS_Psi_Inicial}
\end{equation}
with $a\in [0, 1]$ and $\phi \in [0, 2\pi)$.

\subsubsection{Unitary evolution}
If the driven battery is perfectly isolated, the the state at time $t$ is found to be

\begin{eqnarray}
    |\psi(t)\rangle &=& \left[ \sqrt{a} \cos\bigl(\theta (t)\bigr)  
     + i \sqrt{1-a}e^{i\phi}\sin\bigl(\theta (t)\bigr) \right]|0\rangle  \nonumber \\
        &+& \left[ i\sqrt{a}\sin\bigl(\theta (t)\bigr) + \sqrt{1-a}e^{i\phi}\cos\bigl(\theta (t)\bigr) \right]e^{-i\Delta t}|1\rangle,  \nonumber \label{eqn:2LS_Psi_Schrodinger}
\end{eqnarray}
where
$  \theta (t) = (g/2) \int_{0}^{t} dt' f(t')
$ is related to the driving function $f(t)$ which, following~\cite{Gemme}, we take 

\begin{equation}
    f(t) = \mathcal{A}\, \exp\left({-\frac{(t-\tau/2)^2}{2\,\sigma ^2}}\right),
    \label{eqn:2LS_Funcion_f}
\end{equation}
with effective amplitude $\tilde g = g\,\mathcal{A} = \theta_m/\big(\sigma\sqrt{\pi/2}\big)$, $\theta_m$ the maximum value achieved by $\theta (t)$ and $\sigma$ the standard deviation.
To ensure a perfect population transition $\sigma \ll \tau$ is required. In this regime, $\theta(t)$ can be approximated by an Error function
\begin{equation}
    \theta (t) \approx  \frac{\theta_m}{2} \left [ \text{Erf} \left ( \frac{t - \frac{\tau}{2}}{\sqrt{2}\sigma}\right ) + 1 \right ].
    \label{eqn:2LS_Theta_aprox}
\end{equation}

Using the expression obtained for the state of the system and the definition in Eq.~\eqref{eq:GP},  we compute the GP accumulated up to time $t$, which reads

\begin{align}
\Phi_g^{u,(2)}(t, a, \phi) = &\arg \Bigl \{ \bigl [ a + (1-a)e^{-i\Delta t} \bigr ]\,\cos(\theta) 
\\
&+ i\bigl [ e^{i\phi} + e^{-i(\Delta t + \phi)} \bigr ]\sqrt{a\,(1-a)}\sin(\theta) \Bigr \} \nonumber \\ \nonumber
&- \sqrt{a\,(1-a)}\, \cos(\phi)\,\theta(t) + \Delta\,(1-a)\,t 
\\\nonumber
&+ \int_{0}^{t}dt' \mathcal{E}^{(2)}_\B(t',a, \phi)
\label{eqn:2LS_GP_Analitica}
\end{align}
with $\mathcal{E}_\B^{(2)}(t,a,\phi)$ the energy stored in the battery.

If $a=1$, the accumulated GP takes the expression

\begin{equation}
    \Phi_g^{u,(2)}(t, 1) = I_{\mathcal{E}^{(2)}}(t),
    \label{eqn:2LS_GP_Analitica_a=1}
\end{equation}
with $I_{\mathcal{E}^{(2)}}(t) \coloneq\int_{0}^{t} dt' \mathcal{E}^{(2)}_\B(t',1,\phi)$,
establishing a direct relationship between the GP and the energy stored in the battery. Moreover, all dependence in the initial relative phase $\phi$ is lost.
Therefore, we can stress that {\it the GP accumulated under unitary evolution up to time $t$, equals the time integral of the stored energy if the initial state is the ground state.}
It is important to state that this relation emerges from the vanishing of several contributions. For instance, the state of the battery remains at all times in phase (in Pancharatnam sense) with the initial state, meaning the scalar product of these states is a positive real number. Moreover, the dynamical phase associated with the internal dynamics of the qubit and the driving vanish as well.

\subsubsection{Non-unitary evolution}

Realistic QBs are experimentally realized in the presence of an external environment, resulting in decoherence and the depletion of quantum resources. The implementation of a quantum battery in practice must address the challenge of environmental interactions. Therefore, we will consider an OQB subjected to a non-unitary evolution, and refer to the unitary evolution as a benchmark. The non-unitary evolution is then ruled by Eq. (\ref{eq:Lindblad}), with the relaxation operator $L_{{\rm r}} = |0\rangle \langle 1|$ and the dephasing operator $L_{{\rm d}} = |1\rangle \langle 1|$ associated with transition rates $\gamma$ and $\eta$, respectively.
To address the effect of the environment on the charging protocol, we restrict ourselves to the case of a perfect initial ground state $a = 1$, and numerically solve the evolution of the density matrix. Following the definition given in Eq.~(\ref{eqn:E_Almacenada_Definicion}), the energy stored in an open evolution can be written in terms of the populations as
\begin{equation}
    \mathcal{E}^{(2)}_{OQB}(t) = \Delta\,\rho_{11}(t),
    \label{eq:2LS_energia}
\end{equation}
and is expected to be different from the value acquired in a unitary evolution.

\begin{figure}[ht]
    \centering            
    \includegraphics[width=0.9\linewidth]{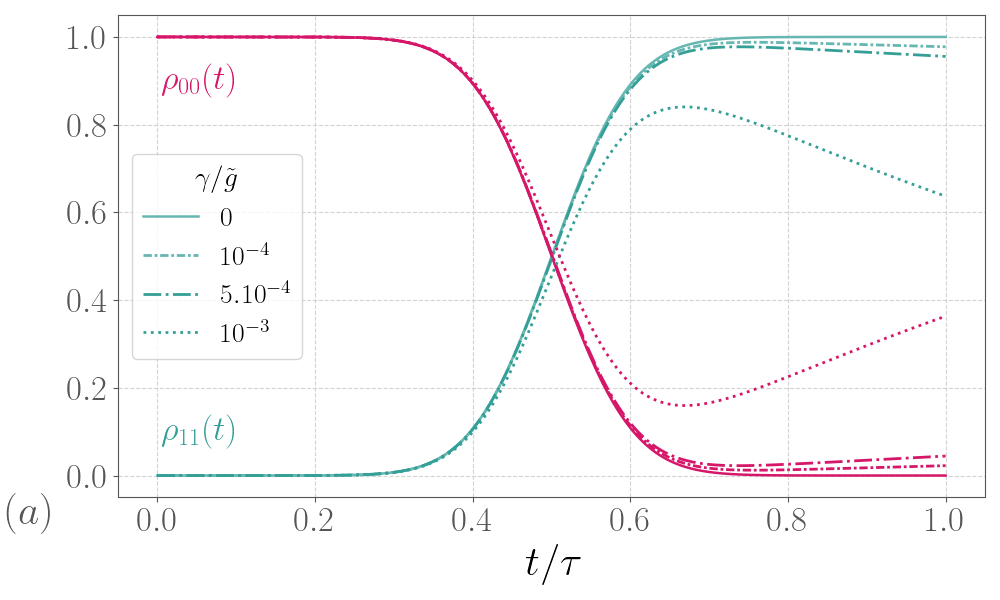}
    \includegraphics[width=0.9\linewidth]{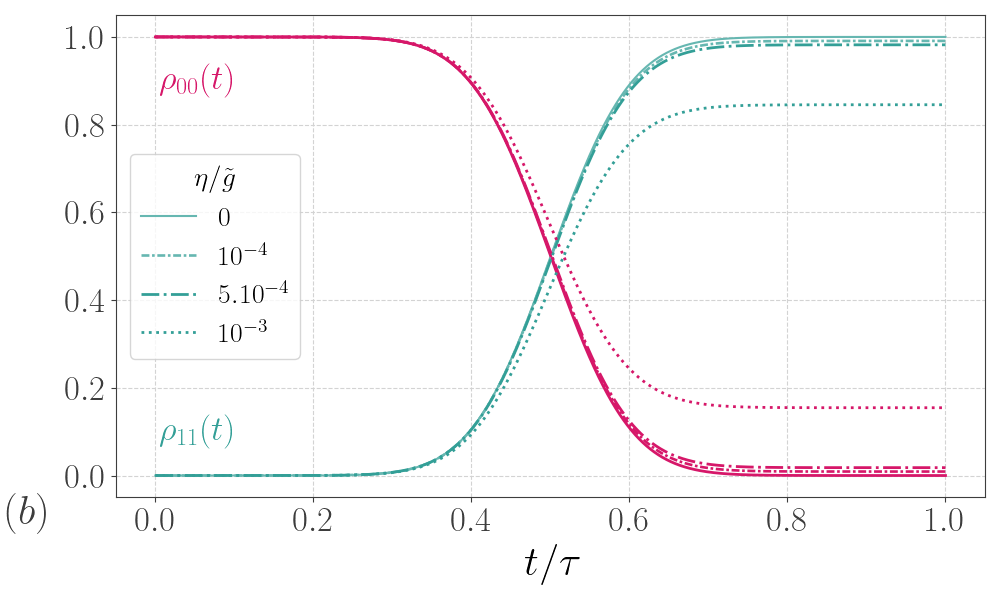}
    \caption{
     Populations along the charging process for different (a) relaxation rates $\gamma/\tilde g$, with $\eta = \gamma/10$, and (b) dephasing rates $\eta/\tilde g$, with $\gamma = \eta/10$. We consider an artificial atom with energy gap $\Delta/\tilde g = 10$,  which is initialized in an initial state defined by $a=1$ and $\phi=0$ and further driven as modeled with parameters $\sigma=\tau/8$ and $\theta_m = \pi/2$.}
    \label{fig:2LS_NonUnitary_CASO}
\end{figure}

As it depends only on the path traced on the physical states space, the GP is related to the behavior of the density matrix. Hence, in order to gain some intuition, we first observe the population dynamics. We split the analysis into a first scenario where the relaxation effect are assumed dominant over dephasing (panel (a) in Fig.~\ref{fig:2LS_NonUnitary_CASO}), and a second setup in which the dominant effect is dephasing (panel (b) in Fig.~\Ref{fig:2LS_NonUnitary_CASO}). We also include the unitary case for reference.

Comparing both panels, several coincidences arise. Although the charging process is, strictly speaking, always affected by the influence of the environment, there are some regimes of the parameters in which the effect of the environment can be neglected.
In these cases, the charging process reaches the maximum value similar to the transition that occurs in an isolated QB.
However, within the regime of parameters in which the environmental effects are non-negligible ($\gamma/\tilde{g} \geq 10^{-3}$ for the parameter values in the plot), qualitative differences arise. Relaxation effects give rise to an unstable process in which the excited-state population first reaches a maximum and then decays,  leading to a hampered practical application (panel (a) in Fig.~\ref{fig:2LS_NonUnitary_CASO}).  
Meanwhile, dephasing environments preserve the qualitative behavior of the evolution, so that the state of the system converges to a fixed point along the vertical axis of the Bloch sphere, corresponding to a stable population distribution.
However, the population distribution differs from the unitary result, as the state is now a mixed state instead of pure excited state (panel (b) in Fig.~\ref{fig:2LS_NonUnitary_CASO}).

\begin{figure}[ht]
    \centering            
    \includegraphics[width=0.95\linewidth]{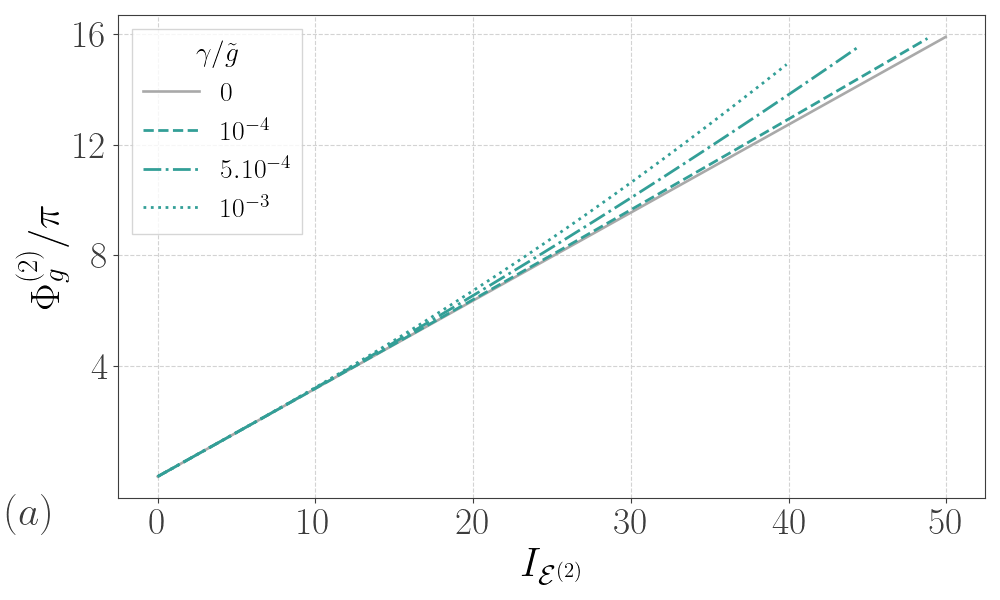}
    \includegraphics[width=0.95\linewidth]{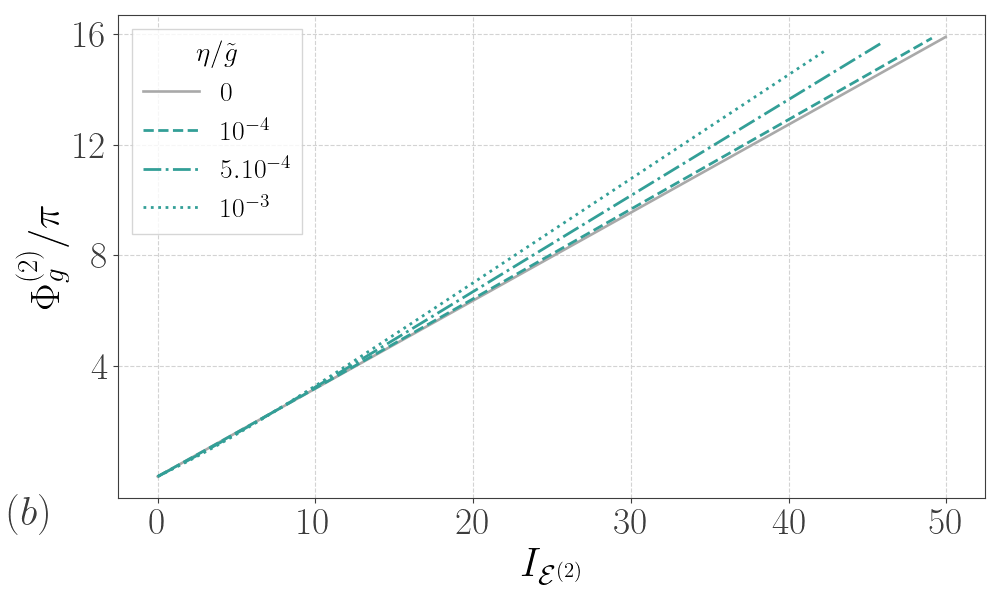}
    \caption{
     Accumulated GP as function of the time-integrated stored energy $I_{\mathcal{E}^{(2)}}$ during the charging process of an OQB. We consider: (a)  relaxation effects  and  (b)  dephasing effects. The unitary case is presented for reference with a gray line. Other parameters as in Fig. \ref{fig:2LS_NonUnitary_CASO}.}
    \label{fig:2LS_fasevsAEP}
\end{figure}

In the following, we numerically compute the accumulated GP for a non-unitary charging process and see what the effect of the environment in the relation derived in Eq.~\eqref{eqn:2LS_GP_Analitica_a=1}. 
In the general case, the environment will deviate both, $\mathcal{E}^{(2)}(t)$ and the GP, from the unitary results, so that the linear relation can not be granted.
Fig.~\ref{fig:2LS_fasevsAEP}, shows the GP as a function of the stored energy integral for different values of the (adimensional) relaxation $\gamma/\tilde g$ (panel (a)) and dephasing $\eta/\tilde g$ rates (panel (b)). 

As previously seen, relaxation effects affect population stability, resulting therefore the most detrimental noise source. For values of the relaxation rate $\gamma/\tilde{g}> 5\;10^{-4}$, the acquired GP significantly deviates from the unitary value $\Phi_g^{u,(2)}$, implying that environmental effects are strongly modifying the path traversed in physical states space.
However, the unitary relation is retained up to rates  $\gamma/\tilde{g} \sim 10^{-4}$,  which are in agreement with the experimental values reported~\cite{PRA107,Song2017}. Panel (b) shows that a similar behavior takes place for dominant dephasing channel. Therefore, for experimental values of $\gamma/\tilde{g}$ and $\eta/\tilde{g} \sim 10^{-4}$ the linear relation between the phase and the energy integral in Eq.~\eqref{eqn:2LS_GP_Analitica_a=1} remains resilient~\cite{Peterer}. For this case of a 2-level QB, the relation between the GP and the stored energy is robust up to considerable realistic values modeling relaxation and dephasing effects.
 Once again, qualitative differences arise in when the environmental effects can not be disregarded. The population instability displayed in relaxation scenarios (panel (a) in Fig.\ref{fig:2LS_NonUnitary_CASO}) discussed above, implies the stored energy is non-monotone on time. Therefore, it's integral increasingly deviates after the maximum. On the other hand, as the environmentally affected system is not approximating a fixed point in the y-axis of the Bloch sphere, its state keeps accumulating non bigger contributions to the GP. The net effect of relaxation effects is therefore the bending of $\phi_g^{(2)}$ as a function of $I_{\mathcal{E}^{(2)}}$ (see Fig.\ref{fig:2LS_fasevsAEP}.a). Differently, in presence of dephasing effects, panel (b) in Fig.\ref{fig:2LS_fasevsAEP}) shows that after an initial regime in which $\phi_g^{(2)}$ is a non-linear function of $I_{\mathcal{E}^{(2)}}$, the linearity is recovered, though with a different slope.

\section{Three level state quantum battery}
\label{3L}
The model studied in Sec. \ref{2L} above has been  generalized by adding an extra energy level to the QB (see Fig. \ref{fig:circuito}). The Hamiltonian defined in Eq. (\ref{QB}) specified for a three-level system can be written as
\begin{equation}  H\B^{(3)} = \omega_0\,|0\rangle \langle0|  + \omega_1\,|1\rangle \langle 1| + \omega_2\,|2\rangle \langle 2|,
  \label{eqn:3LS_H_Q}
\end{equation}
where $\{|0\rangle,|1\rangle,|2\rangle\}$ are the different energy eigenstates. Herein, we shall define  $\Delta = \omega_1 - \omega_0$ and $\Delta ' = \omega_2 - \omega_1$, as the energy gaps between consecutive levels.  Experimentally, the implementation of a quantum battery in a transmon \cite{santos2019stable, hu2022optimal, Dou2018,Dou2022,Gemme} implies $\Delta \neq \Delta'$ due to the anharmonicity exhibited by these systems since $\Delta= \omega_p - E_c$ and $\Delta'=\Delta- E_c$ (see Fig. \ref{fig:circuito}).
Once again, we shall consider the charger as a classical device and, therefore, we do not address its dynamics. The time dependent Hamiltonian is
\begin{equation}
  H_o^{(3)}(t) = H_\B^{(3)} + \hat{V}^{(3)}(t),
  \label{eqn:3LS_H_tot}
\end{equation}
where $\hat{V}^{(3)}$ is an external driving field, with $t\in[0,\tau]$, that couples consecutive energy levels as given by
\begin{eqnarray}
  \hat{V}^{(3)}(t) &=& g^{(3)}\,f_{01}(t)\cos(\Omega_{01}\, t) \Bigl(|0\rangle \langle 1| + |1\rangle \langle 0| \Bigr) \nonumber \\
  &+& g^{(3)}\,f_{12}(t)\cos(\Omega_{12}\, t) \Bigl( |1\rangle \langle 2| + |2\rangle \langle 1| \Bigr),
  \label{3LS:H_int}
\end{eqnarray}
with  $f_{01}(t)$ y $f_{12}(t)$ arbitrary coupling functions. Herein, we choose $f_{01}(t)=f_{12}(t)=f(t)$, with $f(t)$ given by Eq.~(\ref{eqn:2LS_Funcion_f})  which implies a simultaneous charging protocol~\cite{Gemme}.
Meanwhile, the most general initial state of the QB can be written as
\begin{equation}
    |\psi (0) \rangle = \sqrt{a}\,|0\rangle + \sqrt{1-a-c}\,e^{i \phi}|1\rangle  + \sqrt{c}\,e^{i \varphi}|2\rangle
    \label{eqn:3LS_PsiInicial}
\end{equation}
where $a, c \in [0, 1]$ and $\phi, \varphi \in [0, 2\pi]$.

\subsubsection{Unitary evolution}
The state of the system under unitary evolution can be exactly computed as

\begin{equation*}
    \!|\psi(t)\rangle = C_0(t)|0\rangle + C_1(t)e^{-i\Delta t}|1\rangle + C_2(t)e^{-i(\Delta + \Delta ') t}|2\rangle,
\end{equation*}
with
\begin{align*}
    C_0(t) &= \frac{1}{2} \left [ \sqrt{a}(\cos(\Theta) + 1)  - i \sqrt{2}\sqrt{1-a-c}e^{i \phi}\sin(\Theta) \right ]\nonumber \\
    &+ \frac{1}{2} \sqrt{c}e^{i \varphi}(\cos(\Theta) - 1),  \nonumber \\
    C_1(t) &= \frac{1}{\sqrt{2}} \left [- i\sqrt{a}\sin(\Theta) + \sqrt{2}\sqrt{1-a-c}e^{i \phi}\cos(\Theta)\right ] \nonumber \\
    &-i\frac{1}{\sqrt{2}}\sqrt{c}e^{i \varphi}\sin(\Theta),   \nonumber \\
    C_2(t) &= \frac{1}{2} \left[ \sqrt{a}(\cos(\Theta) - 1)
    - i \sqrt{2}\sqrt{1-a-c}e^{i \phi}\sin(\Theta) \right]\nonumber \\
    &+ \frac{1}{2}\sqrt{c}e^{i \varphi}(\cos(\Theta) + 1),  \nonumber 
\end{align*}
where $\Theta \equiv \Theta\! (t)=\!(g^{(3)}/\sqrt{2})\int_{0}^{t} dt' f(t')$, which can be approximated as
\begin{equation}
    \Theta (t) \approx  \frac{\Theta_m}{2} \left [ \text{Erf} \left ( \frac{t - \frac{\tau}{2}}{\sqrt{2}\sigma}\right ) + 1 \right ].
    \label{eqn:3LS_Theta_aprox}
\end{equation}
We introduce the 3-level effective amplitude $\tilde{g}^{(3)} = \Theta_m/(\sigma\sqrt{\pi})$. 
Once again, the success of a perfect transition $|0\rangle \rightarrow |2\rangle$ during the protocol lies in the correct preparation of the initial state in the ground state ($a=1$) in as much as in the subsequent evolution.  As we have done with the 2-level quantum battery, we shall consider the charging process from perfect ground state preparation occurring in an open way, and refer to the unitary case as a reference. 

Using the expression for state at time $t$, we compute the GP acquired during a unitary evolution

\begin{widetext}
\begin{align}\label{eqn:3LS_GP_Analitica}
    \Phi_g^{u,(3)}(t, a, c, \phi, \varphi) &= \arg \left \{  \sqrt{a}\;C_0(t) + \sqrt{1-a-c}\;e^{-i\Delta t-i\phi}\,C_1(t) + \sqrt{c}\;e^{-i(\Delta' + \Delta)t-i\varphi}\,C_2(t) \right \}\\\nonumber
    & + \sqrt{1-a-c}\;\Theta(t)\big[\sqrt{2\,a}\cos(\phi)+ \sqrt{2\,c}\cos(\phi-\varphi)\big] + (1-a-c)\,\Delta\,t + c\,(\Delta' + \Delta)\,t\\\nonumber
    & + \! \int_{0}^{t}\!\!dt'\; \mathcal{E}^{(3)}_\B(t', a, c, \phi, \varphi).
\end{align}
\end{widetext}

The GP in Eq.\eqref{eqn:3LS_GP_Analitica} above depends on the preparation of the initial state, described by the parameters $a,c,\phi,\varphi$, on the transition energies $\Delta, \Delta'$ and the characterization of the driving external field $\Theta(t)$. 

However, similarly to the case of the two-level QB, for $a=1$ y $c=0$,  all dependence in the relative phases $\phi$ and $\varphi$ is lost, and the accumulated GP reduces to the time integral of the stored energy  $\mathcal{E}^{(3)}(t,1,0,\phi, \varphi)$,
 
\begin{equation}
    \Phi_g^{u,(3)}(t,1,0) = I_{\mathcal{E}^{(3)}}(t)
    \label{eqn:3LS_GP_Analitica-a=1},
\end{equation}
with $I_{\mathcal{E}^{(3)}}(t) \coloneq\int_{0}^{t}\,dt'\, \mathcal{E}_\B^{(3)}(t',1,0,\phi,\varphi)$,
establishing, once again, a simple relation among the two quantities: the unitary GP of the three-level system equals the time integral of the unitary energy stored in the $|0\rangle \rightarrow |2\rangle$ transition.
In the following, we shall see to what extent the environment deviations $\delta \Phi$ affect this relationship.

\subsubsection{Non-unitary evolution}
\begin{figure}[h]
    \centering               
    \includegraphics[width=0.95\linewidth]{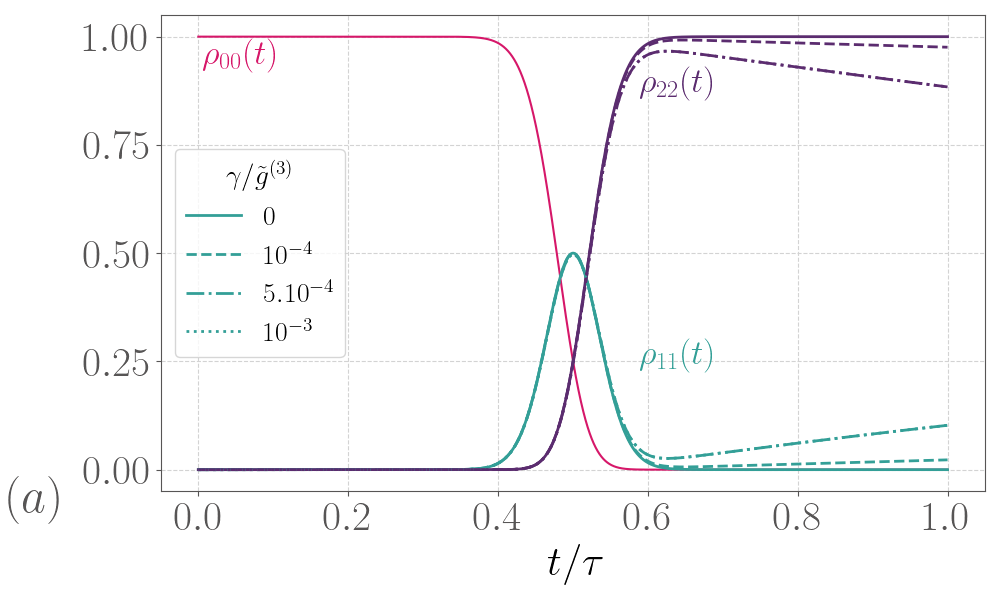}
   \includegraphics[width=0.95\linewidth]{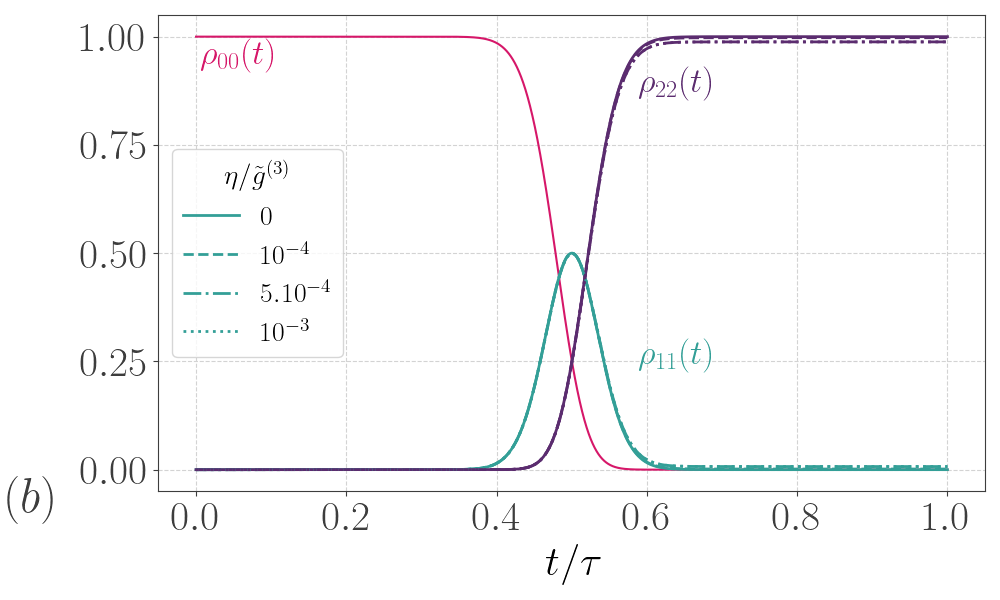}
    \caption{
    Population dynamics during the charging process of the three-level OQB for different (a) values of dominant $\gamma_{10}$, and (b) values of dominant $\eta_{1}$. We consider an artificial atom with energy gaps $\Delta/\tilde{g}^{(3)}=100$, $\Delta'/\tilde g^{(3)} = 95$, which is initialized in an initial state defined by $a=1, c=0$. It is further driven as modeled by $\sigma=\tau/16$ and $\Theta_m = \pi$. The dissipation and dephasing channels are set $\gamma=\gamma_{21}=\gamma_{10}$, and $\eta=\eta_1= \eta_2$.}
    \label{fig:3LS-poblaciones_NU}
\end{figure}    

In this open 3-level quantum battery, the dynamics is ruled by Eq.(\ref{eq:Lindblad}) with $L_{{\rm r},k} = |k-1\rangle\langle k|$ and with $L_{{\rm d},k} = |k\rangle\langle k|$. Following the experimental values reported \cite{Peterer}, we assume the particular situation of
$\gamma_{21} =\gamma_{10}$ and $\eta_1= \eta_{2}$ in our simulations.  In the same way that we did for the two-level model, we get the first insight through the observation of the state populations displayed in Fig.\ref{fig:3LS-poblaciones_NU}. In panel (a), we show the scenario of dominant relaxation effects during the charging process, while in panel (b), dephasing effects are dominant. 
The dissipative environment in panel (a) shows that the charging process is very unstable after the system has reached the highest energy level. 
However, in the case of dephasing, the OQB appears to be more resistant to noise and provides a fully stable charge for corresponding values of $\eta$. This is a considerable effect of the environment on a qutrit and an advantage with respect to the above two-level OQB under dephasing.
In fact, qubit's relaxation has been reported to play an inhibition role in the stable charging process of a quantum battery~\cite{Dou2022}.

The energy stored during the charging process Eq.~(\ref{eqn:E_Almacenada_Definicion}) can be expressed in terms of the diagonal elements of the density matrix as
\begin{equation}
    \mathcal{E}^{(3)}_{OQB}(t) = \Delta\,\rho_{11}(t) + (\Delta + \Delta')\,\rho_{22}(t).
\end{equation}

\begin{figure}[h]
    \centering               
    \includegraphics[width=0.95\linewidth]{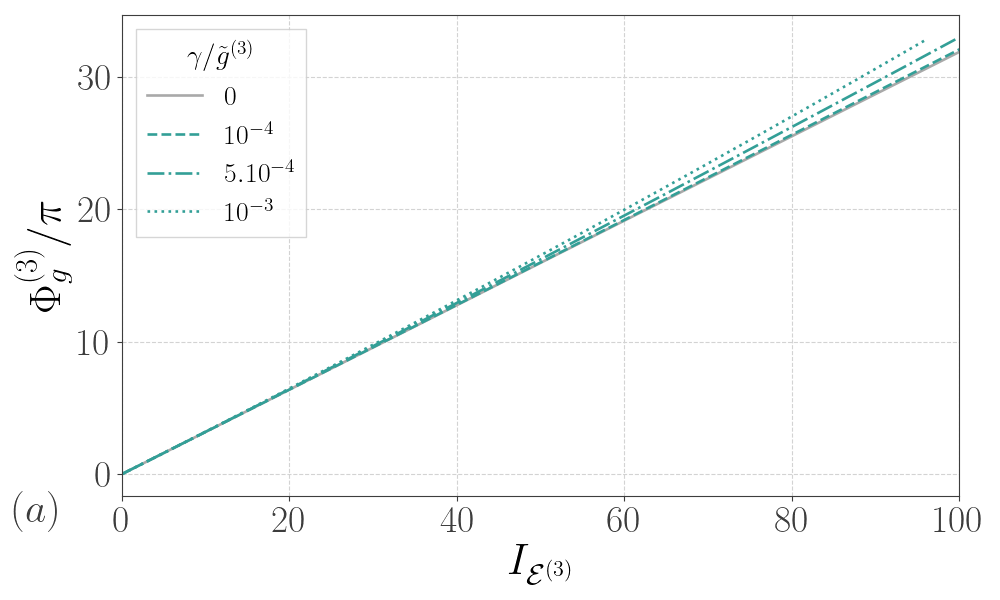} \includegraphics[width=0.95\linewidth]{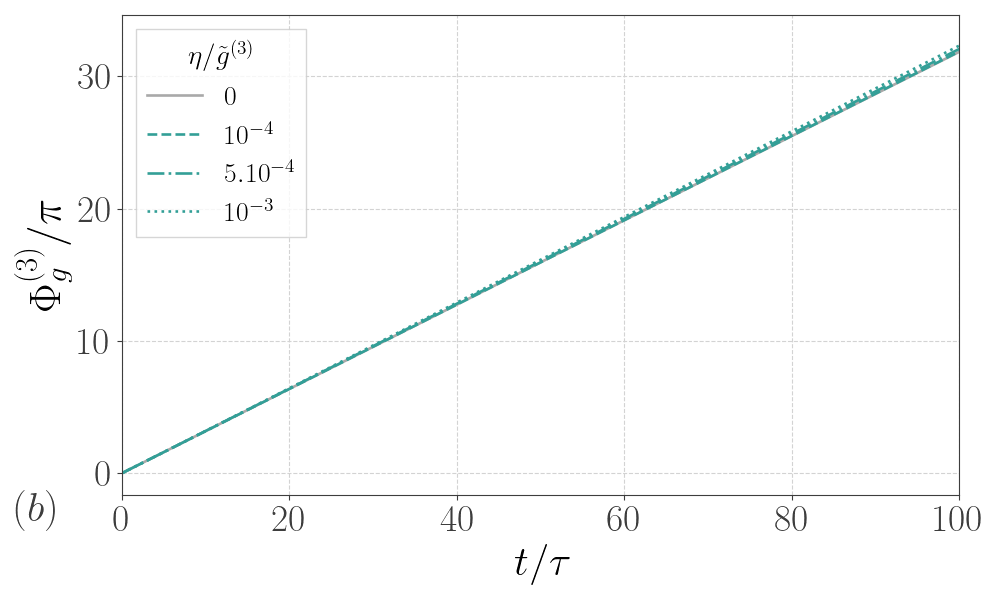}
    \caption{GP accumulated during the charging process for a three level system as function of the stored energy integral $I_{\mathcal{E}^{(3)}}$ for dominant relaxation (panel (a)) and dominant dephasing effects (panel (b)). Parameters as in Fig. \ref{fig:3LS-poblaciones_NU}.}
 \label{fig:3LS-fasevsEnergia}
 \end{figure}

Fig.~\ref{fig:3LS-fasevsEnergia} shows the results for a charging process implemented in a three-level system with a driving force characterized by Eq.~(\ref{3LS:H_int}). Specifically,  in panel (a) of Fig.~\ref{fig:3LS-fasevsEnergia}, we present the results for a non-unitary charging process with a dominant dissipation channel characterized by the relaxation rate $\gamma$, while in panel (b),  the  charging process is done under a dominant dephasing channel characterized by $\eta$. In both cases, in the limit of no environment ($\gamma/\tilde g^{(3)}\rightarrow 0$ and $\eta/\tilde{g}^{(3)} \rightarrow 0$), the relation between the accumulated GP and the  time-integrated stored energy remains linear as the analytic expression of Eq.~(\ref{eqn:3LS_GP_Analitica-a=1}) predicts, say the environmental corrections are zero. For small enough dephasing effects ( $\eta/\tilde{g}^{(3)} \sim 10^{-3}$), the relation still holds even though the evolution is open. On the other hand, in the relaxation channel, the behavior is somewhat different: smaller values of $\gamma/\tilde{g}^{(3)}$ are needed to retain the validity of the proportional relation compared to the dephasing channel. 

\begin{figure}[h]
    \centering         
    \includegraphics[width=0.95\linewidth]{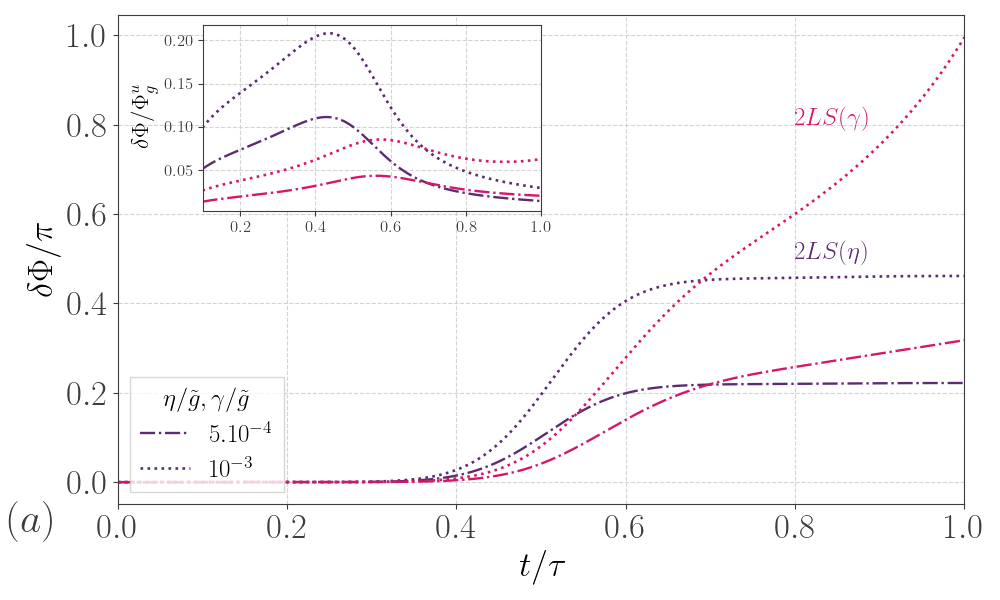} \includegraphics[width=0.95\linewidth]{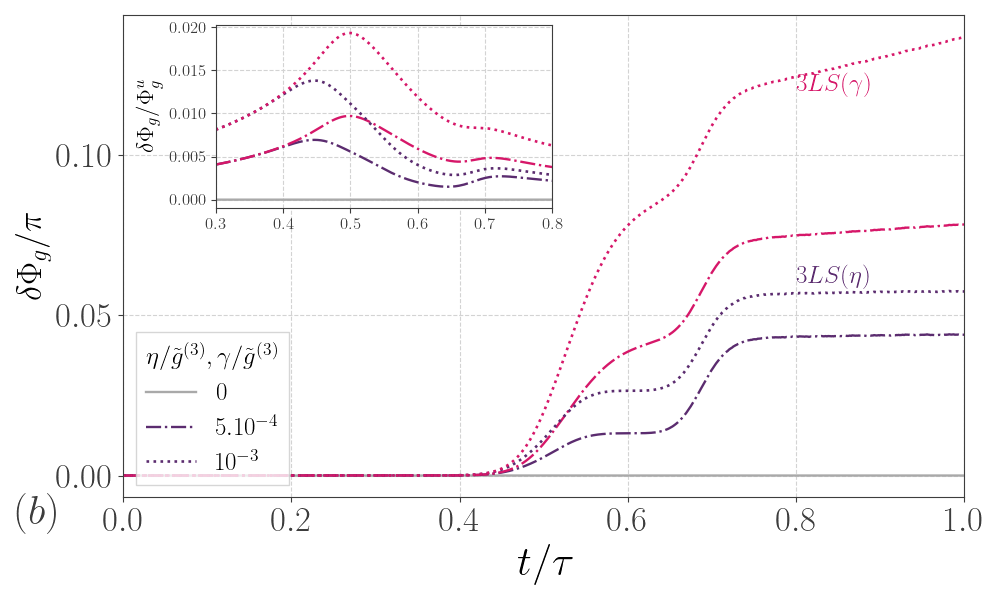}
    \caption{ Environmental corrections $\delta \Phi_g/\pi$ induced on the GP under relaxation and dephasing channels when: panel (a) 2LS is considered and panel (b) a 3LS is considered. Parameters as in Fig. \ref{fig:3LS-poblaciones_NU}.}
    \label{fig:comparacion}
\end{figure}

Fig.~\ref{fig:comparacion} compares the deviations induced in a 2L open battery (panel (a)) and a 3L open battery (panel (b)) which are charged under relaxation effects and dephasing channels. In both batteries, dephasing channel yields a smaller correction from the unitary GP at the end of the charging process. 
By simultaneously observing panel (a) and (b), we note that the deviations $\delta \Phi_g$ are consistently smaller in the 3LS than in the 2LS, with relative deviations differing by an order of magnitude in each case (see insets). This robustness of the GP for the 3LS battery combined with the robustness observed in the populations in Fig.\ref{fig:3LS-poblaciones_NU}, yields the extended validity of the linear relation Eq.~\eqref{eqn:3LS_GP_Analitica-a=1} which is observed in Fig.\ref{fig:3LS-fasevsEnergia}.
Therefore, for 2LS and 3LS batteries which are driven such that the charging time $\tau$ remains the same in both cases, the 3LS battery shows an extended robustness which agrees with the results reported for a larger Hilbert space such as a bipartite system in a dephasing environment \cite{bipartite, Lucho, Soba}.
In \cite{Zhang2018, Zhang2019}, authors reported the experimental observation of the GP in a qutrit coupled to a resonator. 
Therefore, the GP obtained during the charging process for an initial ground state is resistant to dephasing noise in a wide range of parameters (in agreement with those reported in cQED) and  directly related to the energy stored by the system.

\section{Quantum Charger Model}
\label{charger}
In this Section, we shall study a QB for which the charger is also considered as a quantum device. 
Recently, QBs models build up from Dicke model~\cite{Andolina2018} were proposed. The simplest, yet nontrivial, case of a charger-battery setting is that where the charger and QB are resonant 2LSs. However, it is important to mention that there are more complex systems where the QB has been experimentally studied as multiple transmon qubits coupled to one mode of a resonator (charger)~\cite{PRA107}. 

\begin{figure}[th]
 \begin{center}
\includegraphics[width=0.85\linewidth]{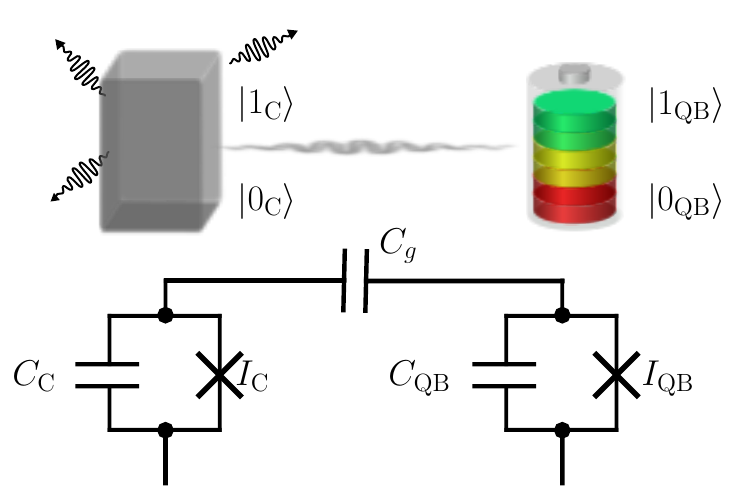}
 \caption{The composite system is formed by a quantum charger and a quantum battery, both considered as two-level systems that interact through a quantum battery-charger coupling. In our particular model, we assume that the charger is driven at a strength $F$ and is additionally subject to dephasing at rate $\eta$.}
 \label{fig:circuitocomposite}
 \end{center}
 \end{figure}

The Hamiltonian of the complete battery-charger system includes contributions: 
\begin{eqnarray}
\hat{H}_o(t)= \hat{H}_{\B} + \hat{H}_{\C} + \hat{H}_{\B,\C} + \hat{V}(t),
\end{eqnarray}
where $\hat{H}_{\B}$ and $\hat{H}_{\C}$ represent internal dynamics of each 2-level subsystem (QB and charger) with corresponding frequency $\omega_\B$ and $\omega_C$

\begin{eqnarray}
\hat{H}_{\C}= \omega_\C\; |1_{\C}\rangle\langle 1_{\C}| \,{\scriptscriptstyle\otimes}\,\mathbb{I}_\B\;\;;\;\;
\hat{H}_{\B} = \mathbb{I}_\C\,{\scriptscriptstyle\otimes}\,\omega_{\B}\; |1_{\B}\rangle\langle 1_{\B}|\nonumber,
\end{eqnarray}
while the other two terms represent interactions: the charger-quantum battery and the external driving acting on a charger

\begin{align}
\hat{H}_{\B,\C} &= g\,|1_\C\,0_\B\rangle\langle 0_\C\,1_\B| + {\rm h.c.} \\[.7em] \nonumber
\hat{V}(t) &= F\;|0_{\C}\rangle\langle 1_{\C}|  e^{i \omega_d t} \,{\scriptscriptstyle\otimes}\,\mathbb{I}_\B + {\rm h.c.},
\end{align}
with $F$ and $\omega_d$ denoting the strength and frequency of the charger driving \cite{Ahmadi,Shastri}. When the resonance condition $\omega_{\B} = \omega_\C$ is satisfied, the battery-charger coupling $\hat{H}_{\B,\C}$ commutes with the bare contributions $H_\B$ and $H_\C$, ensuring that there is no energetic cost to switch on/off the interaction (in the absence of driving). {We address the case with $\omega_d = \omega_\B = \omega_\C$.}
Under unitary conditions energy will be transferred between the charger and the QB due to their interaction, making it impossible to reach a stable storage energy~\cite{Andolina2018}.
However, dissipation effects modify this situation, and it has been recently shown that moderately dephased charger leads to efficient charging~\cite{Shastri}.

\begin{figure}[ht]
 \begin{center}
 \includegraphics[width=0.85\linewidth]{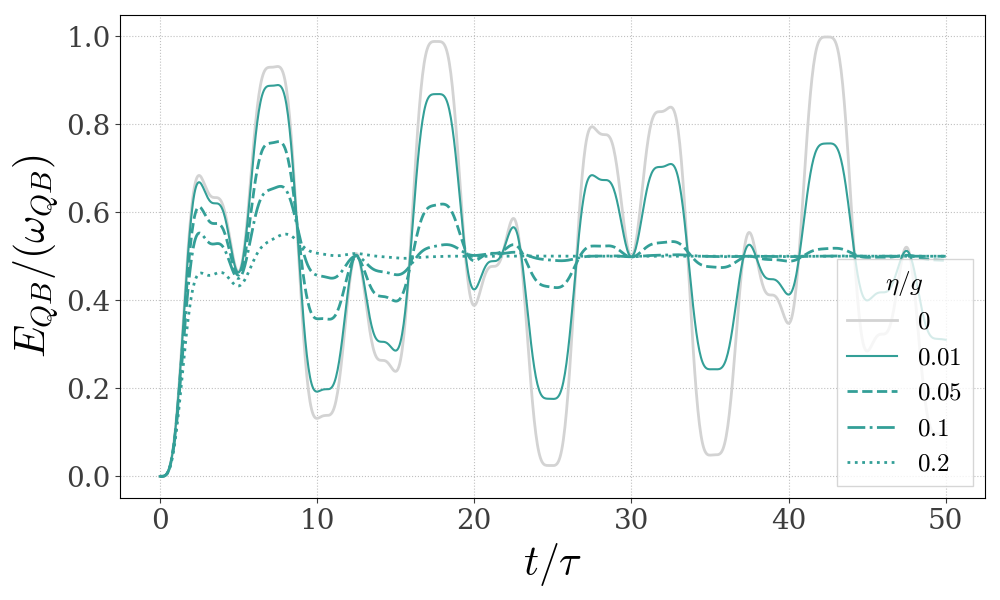}
 \caption{Energy stored in the QB after the coherent driving of the charger for different values of dephasing factor $\eta/g$. The gray line represents the energy stored under an unitary evolution. Color lines represent different values of dephasing as indicated in the legend. Parameters used:  for the classic driving we consider $F/g=0.5$, $\tau = 2\pi/\omega_\B$. The values of the dephasing rate and coupling constant are selected following \cite{Shastri} for an optimal charging using a dephased charger.  }
 \label{fig:charger}
 \end{center}
 \end{figure}

Therefore, in the following, we restrict ourselves to the case in which the coupling to the environment is through the charger, which is affected only by dephasing effects.
Therefore, the dynamics of the composite system is described by the master equation in Eq.~\eqref{eq:Lindblad} with Lindblad operators $L_{{\rm d}, \C} = |1_\C\rangle\langle 1_\C|\,{\scriptscriptstyle \otimes}\,\mathbb{I}_{\B}$. Thus, we assume that the charger is driven at a strength $F$ and is additionally subject to dephasing at the rate $\eta$. We consider the initial state to be the bare ground state for both the holder and the charger $|\Psi(0) \rangle = |0_\C\,0_{\B}\rangle$.
Fig.~\ref{fig:charger} presents the energy stored by the QB. As before, 
$E(t)=\rm Tr[\rho_\B(t) \hat{H}]$ with $\rho_\B(t)=\rm Tr_\C[\rho(t)]$.
The energy accumulated at time $t$ is then ${\cal E}_\B(t)=E_\B(t)-E_\B(0)$. 
When there is no dephasing channel ($\eta=0$), the energy stored oscillates as described above, without reaching a steady value (gray line). However, when the dephasing rate is non-vanishing, these oscillations are damped, and the stored energy reaches a steady value. The higher the dephasing rate, the sooner the state relaxes and becomes stable. In this way, the environment becomes a resource, as the energy oscillations of the closed system do not allow for any storage at all. 

All in all, the dynamics can be qualitatively thought as follows: the initial product of the two bare groundstates $|0_\C\,0_\B\rangle$ gets excited due to the external driving acting on the charger and to the charger-battery coupling. Dephasing effects acting on the charger translates in non-decaying populations and damped coherences for the bipartite state. This is, a dephasing environment also for the QB subsystem. We have shown that there exists a relationship between the accumulated GP and the stored energy in a driven 2LS battery that we probe in the following.

\begin{figure}[ht]
 \begin{center}
\includegraphics[width=0.90\linewidth]{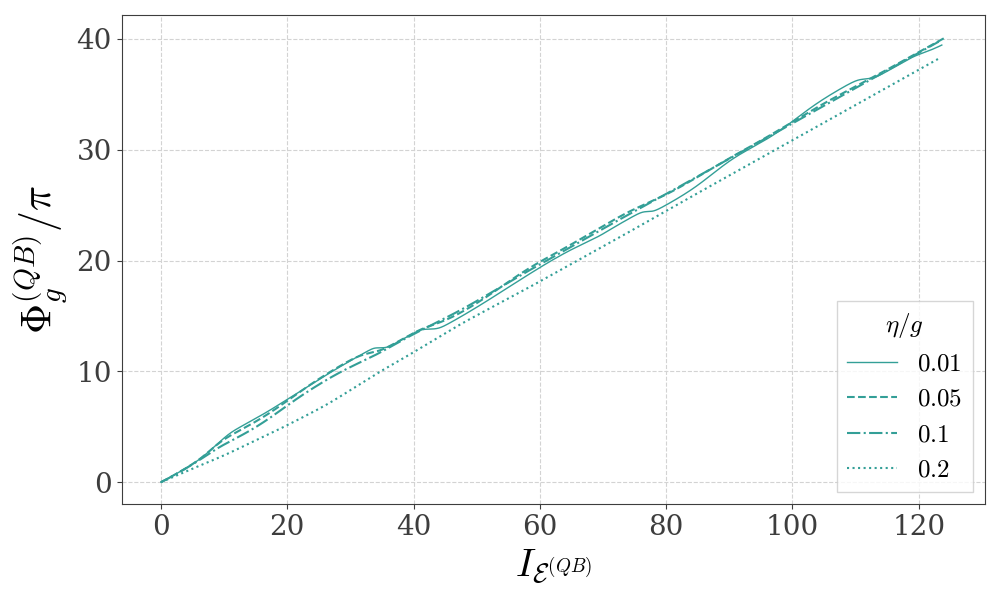}
 \caption{Accumulated GP for the QB as a function of the time-integrated stored energy . The different lines represent the GP acquired during the evolution under the presence of dephased charger.}
 \label{fig:FASE2X2}
 \end{center}
 \end{figure}

In order to compute the  $\Phi_g^{(QB)}$ accumulated by the state of the QB,  we solve  the dynamics of the bipartite system ruled by Eq.~(\ref{eq:Lindblad}). We afterwards trace out the charger and compute the $\Phi_g^{(QB)}$ and the energy accumulated by the QB alone. Fig.\ref{fig:FASE2X2}, shows the accumulated GP as a function of the integral of the stored energy $I_{{\cal E}^{(QB)}}$, with different lines corresponding to different dephasing rates. All cases show a linear relation between the accumulated GP and the $I_{{\cal E}^{(QB)}}$ of the QB (the dotted oscillation corresponds to a value of dephasing where the energy of the QB still oscillates for a while). For values where the energy has already reached a stable value, the relation between the GP and $I_{{\cal E}^{(QB)}}$ seems to present a strong correlation. Remarkably, even though the system evolves under strong dephasing effects, the GP accumulated remains linked to the energy accumulated by the QB. This dissipation mechanism appears to be less detrimental on the GP as some previous results have reported~\cite{Lucho}.

\section{Conclusions}
\label{conclusiones}

 Every quantum system can be considered as an open system because of the unavoidable mutual interaction with an environment. QBs are quantum systems working as energy storage devices, and are therefore susceptible to environmental effects. In this manuscript, we have studied different QBs models. In particular, we have focused on the dynamics of the system under the transition from the ground state to a state of higher energy, driven by an external classical field. These models can be implemented, for example, by quantum devices known as transmons in cQED platforms.

We have first studied the most simple model of a QB considering a 2LS which transitions from the ground state to the first excited state driven by a classical field. During this transition, the system stores energy. 
In the case of an initial ground state and negligible dissipation (unitary evolution), the transition is perfect. If there is dephasing noise in the open evolution, the transition is considerably robust, acquiring a stable final state. However, if relaxation noise is dominant in the open evolution, the final state is not stable. We have used reported experimental values in our simulations. 

We have computed the energy stored in the system and compared it with the accumulated GP. The GP is computed for a non-adiabatic transition under general evolution (i.e. either unitary or non-unitary).
In the case of an unitary evolution, the computation can be done analytically. In this particular case, we have shown that there exists a linear relation between the accumulated GP and the integral of the stored energy. More important, this relation still holds for an open evolution, considering small enough values of noise which are indeed experimentally accessible. For stronger environments, the relationship is broken.

Then, we have extended the analysis to a three-level state QB, which has been reported in IBM platforms and transmon qutrit. 
We have analytically computed the unitary case for reference and provided the accumulated GP. In addition, we have numerically solved the open dynamics by considering relaxation and dephasing channels, representing the first analysis on open three-level QB. We have numerically simulated the sequential protocol by which the ground state is driven to the second excited state. In the case of an initial pure ground state and negligible noise effects, the transition is perfect. However, in the case of an open three-level QB, the system does not reach a stable final state for relaxation dominant effects, whereas for a dominant dephasing channel, the transition to a higher level leads to a stable situation. This means that relaxation noise should be controlled and minimized to a greater extent than dephasing noise for this three-level QB model. 

We have analytically computed the GP for a three-level QB in the limit case of no environment and found the GP equals, for the initial ground state, the time integral of the stored energy. 
We have seen that the relation is preserved along a wider range of parameters compared to the 2LS model. In particular, the GP of a 3LS under dephasing noise is highly resilient and proportional to the accumulated energy stored.

Finally, we have considered a composite QB model composed of a quantum charger and a QB, both described by a 2LS. In particular, we have assumed a dephased charger model. We have seen that, in this particular case, the QB simulates a dephasing evolution, re-obtaining results of the 2LS model. We have numerically computed the accumulated GP for the QB subsystem and compared it with the integral of the energy stored of the QB in a bipartite quantum system. The linear relation appears to hold for different values of the dephasing rate.

Our theoretical findings can be readily verified on state-of-the-art quantum technology platforms. Moreover, key features needed to implement our central idea, namely the ability to control the dephasing strength and to measure GPs, have already been demonstrated in the experimental platforms of superconducting qubits \cite{Berger} and NMR systems \cite{Joshi}.

\section*{Acknowledgments}

This research was funded by Agencia Nacional de Promocion Científica y Tecnológica (ANPCyT), Consejo Nacional de Investigaciones Cientıficas y Técnicas (CONICET), and Universidad de Buenos Aires (UBA). 

\bibliography{referencias}

\end{document}